\documentclass[preprint,showpacs,preprintnumbers,amsmath,amssymb]{revtex4}

\makeatletter
\def\@dotsep{4.5}
\makeatother
\usepackage{graphicx}% Include figure files
\usepackage{dcolumn}% Align table columns on decimal point
\usepackage{bm}% bold math

\begin{document}

%\preprint{Preprint}

\title{Stochastic dynamic model of a growing network based on self-exciting point process}
\author{Michael Golosovsky\footnote{electronic address: michael.golosovsky@mail.huji.ac.il} and Sorin Solomon}
\affiliation{The Racah Institute of Physics, The Hebrew University of Jerusalem, 91904 Jerusalem, Israel\\}
\date{\today} 
\begin{abstract} 
We put under experimental scrutiny the preferential attachment model that is commonly accepted as a generating mechanism of the scale-free complex networks. To this end we chose citation network of Physics papers and traced  citation history of 40,195 papers published in one year. Contrary to common belief, we found that citation dynamics of the individual papers follows the \emph{superlinear} preferential attachment, with the exponent $\alpha= 1.25-1.3$. Moreover, we showed that the citation process cannot be described as a memoryless Markov chain since there is substantial correlation between the present and recent citation rates of a paper.  Basing on our findings  we constructed a stochastic growth model of the citation network, performed numerical simulations based on this model and achieved an excellent agreement with the measured citation distributions.

\end{abstract} 
%L8 Soft Matter, Biological, and Interdisciplinary Physics
\maketitle
%%%%%%%%%%%%%%%%%%%%%%%%%%%%%%%%%%%%%%%%%%%%%%%%%%%%%%%%%%%%%%%%
%The field of complex networks  attracted a  interest in the physics community during last decade 
The field of growing complex networks (informational, social, biological, etc.) attracted increasing interest in the physics community during past decade \cite{Barabasi,Dorogovtsev2002,Newman_SIAM}. Many of these  networks are  believed to achieve  stationary state and to become scale-free \cite{Barabasi,Simon,Solla}.  The static  characteristics of  growing networks such as  clustering coefficient,  community structure, and  degree distribution were extensively studied both theoretically and empirically \cite{Barabasi,Dorogovtsev2002,Newman_SIAM} while the dynamics of these networks  was studied mostly theoretically. It is widely believed that they are generated by the preferential attachment \cite{Barabasi} (cumulative advantage \cite{Solla}) mechanism.  
%Our goal is to put under experimental scrutiny the  underlying assumptions of the preferential attachment mechanism.  We chose citations to scientific papers as one of the best documented networks and a prototype for the study of dynamic behavior of growing networks \cite{Borner}. 
The latter  assumes that new links  are distributed between existing nodes with  probability $\Pi_{i}=\lambda_{i}/\sum_{i}\lambda_{i}$ where
 $\lambda_{i}$ is the attractivity i.e., the expected number of  links acquired by a node $i$ in a short time interval $\Delta t$  \cite{Barabasi}. From the perspective of a single node, the  number of incoming links grows according to the inhomogeneous Markov process  with the  rate 
\begin{equation}
\lambda_{i}=A(k_{i}+k_{0})^{\alpha}
\label{rate}
\end{equation}
where $k_{i}$  is the number of existing links, $k_{0}$ is the  "initial attractivity",  $\alpha$ is the attachment exponent,  $t$ is the age of the node,  and $A(t)$ is the aging function \cite{Dorogovtsev2002,KRL}. In fact, Eq. \ref{rate} describes the stochastic multiplicative growth process 
\begin{equation}
\Delta k_{i}=\lambda_{i}\Delta t+\sigma dW(t)
\label{delta-k}
\end{equation}
where  $\Delta k_{i}$ is the actual  number of newly acquired links during time interval $\Delta t$ and $\sigma dW(t)$ is its stochastic component. 

The direct way to verify Eq.\ref{rate}  is to measure  $\Delta k_{i}$-distributions for the sets of nodes with the same degree  $k$, to  find $\lambda=\overline{\Delta k_{i}}$, and to check how $\lambda$ depends on $k$.  Previous studies that were aimed at this goal \cite{Jeong,Eom,Redner2005,Wang2008}, focussed on the citations to scientific papers as one of the best documented networks and a prototype for the study of dynamic behavior of growing networks \cite{Borner}. Since the above studies were restricted to  relatively small or inhomogeneous  data sets, they had to apply  indirect  averaging procedures, such as numerical integration \cite{Jeong,Eom} or moving average \cite{Redner2005,Wang2008}. These procedures  are prone to quantization errors and yield inconclusive results.
%This  requires a large and homogeneous data set.

Our goal is the direct measurement of the average growth rate of the node degree in a complex network (Eq.\ref{rate}) and the assessment of its stochastic part (Eq.\ref{delta-k}) as well.  Following the accepted practice  \cite{Jeong,Eom,Redner2005,Wang2008} we chose a network of citations to scientific papers. We performed high-statistics and time-resolved study of the citation dynamics of a very large set of papers that is field- and age-homogeneous  (one scientific discipline, one publication year).  Basing on our findings  we constructed a stochastic  model of citation dynamics with no "hidden" parameters such as fitness \cite{Bianconi} or relevance \cite{Medo}. Then we performed numerical simulation based on our model and verified that the real  and simulated  citation networks have the same microscopic and macroscopic characteristics.
%- and compared our results to smaller-scale studies of the Mathematics and Economics papers (SI II)

\begin{figure}
\includegraphics[width=0.45\textwidth]{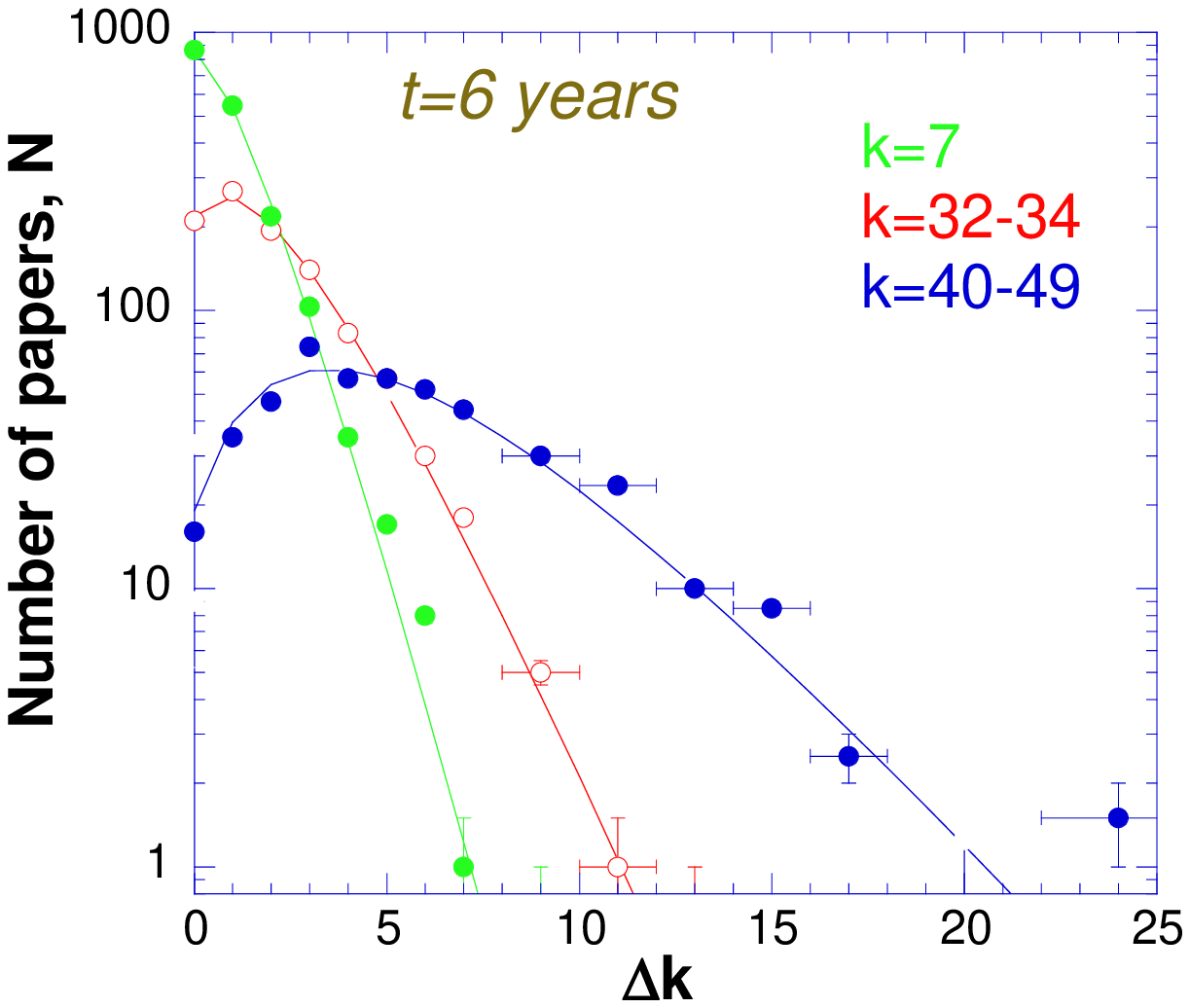}
\includegraphics[width=0.45\textwidth]{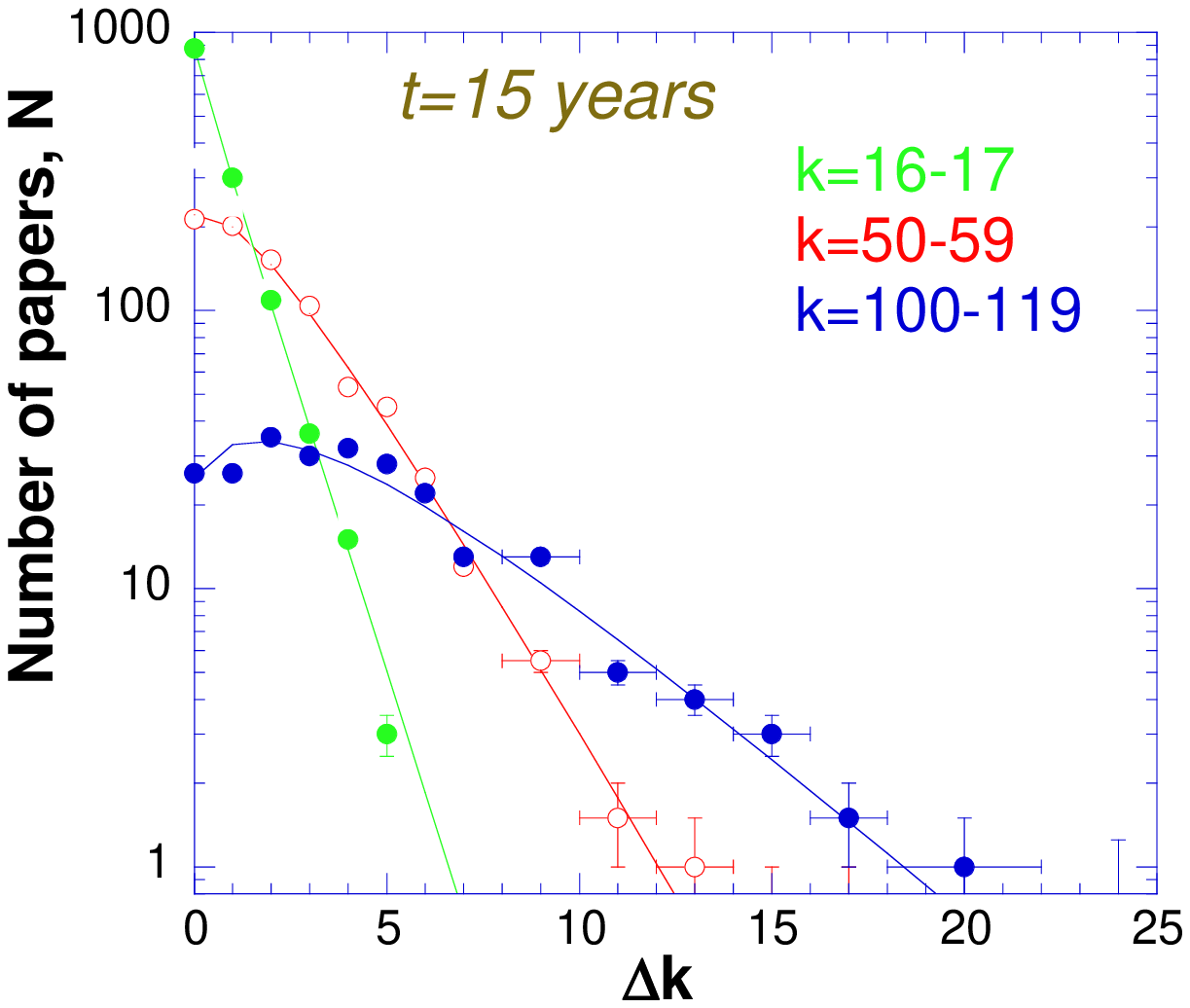}

\caption{Statistical distribution of additional citations $\Delta k_{i}$ accumulated during time window of $\Delta t=1$ year. Continuous lines show fits to negative binomial distribution. $k$ is the number of previous citations and $t$ is the number of years after publication.
}
\label{fig:NG}
\end{figure}

We used  the Thomson-Reuters ISI Web of Science, chose 82 leading Physics journals, excluded review articles, comments, editorial, etc.,  and analyzed  citation history of  all 40,195  original research Physics papers  published in these journals in one year -1984.  For each paper $i$  we determined $k_{i,t}$ - the total number of  citations accumulated  after  $t$ years ($t=T_{cit}-T_{publ}+1$),  and $\Delta k_{i,t}$- the number of  citations gained by the same paper in the  year  $t+1$.  For every citing year $t$ we grouped all papers into $\sim$ 40 logarithmically-spaced bins,  each bin containing the papers with close  $k$.  Figure \ref{fig:NG} shows  statistical distributions  of  $\Delta k_{i}$ for several such bins and for two selected years. For each bin  we found the  mean, $\lambda(k)=\overline{\Delta k_{i}}$, and the variance, $\sigma^{2}=\overline{(\Delta k_{i}-\lambda)^{2}}$. 

\begin{figure}
\includegraphics[width=0.4\textwidth]{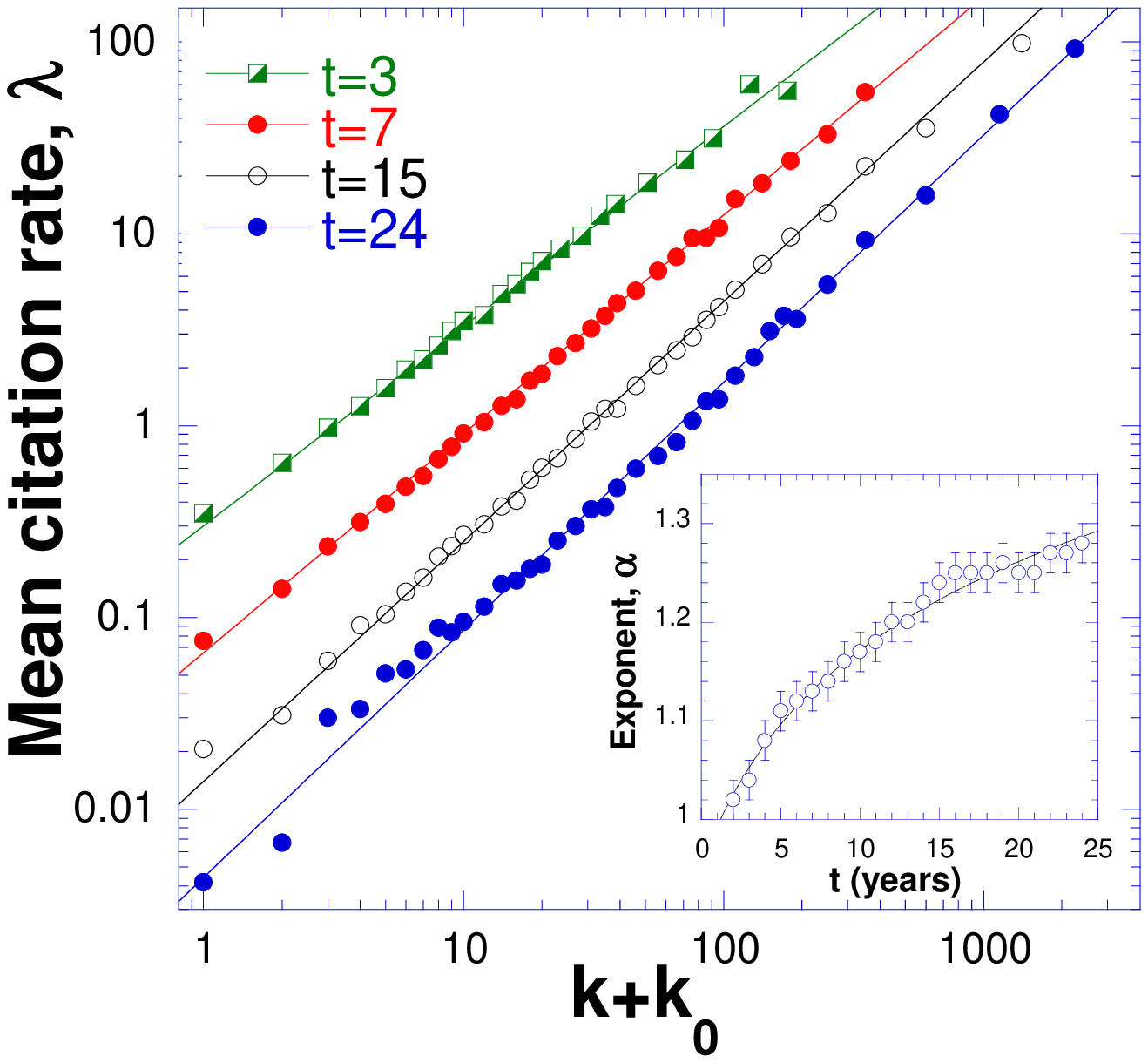}
\includegraphics[width=0.4\textwidth]{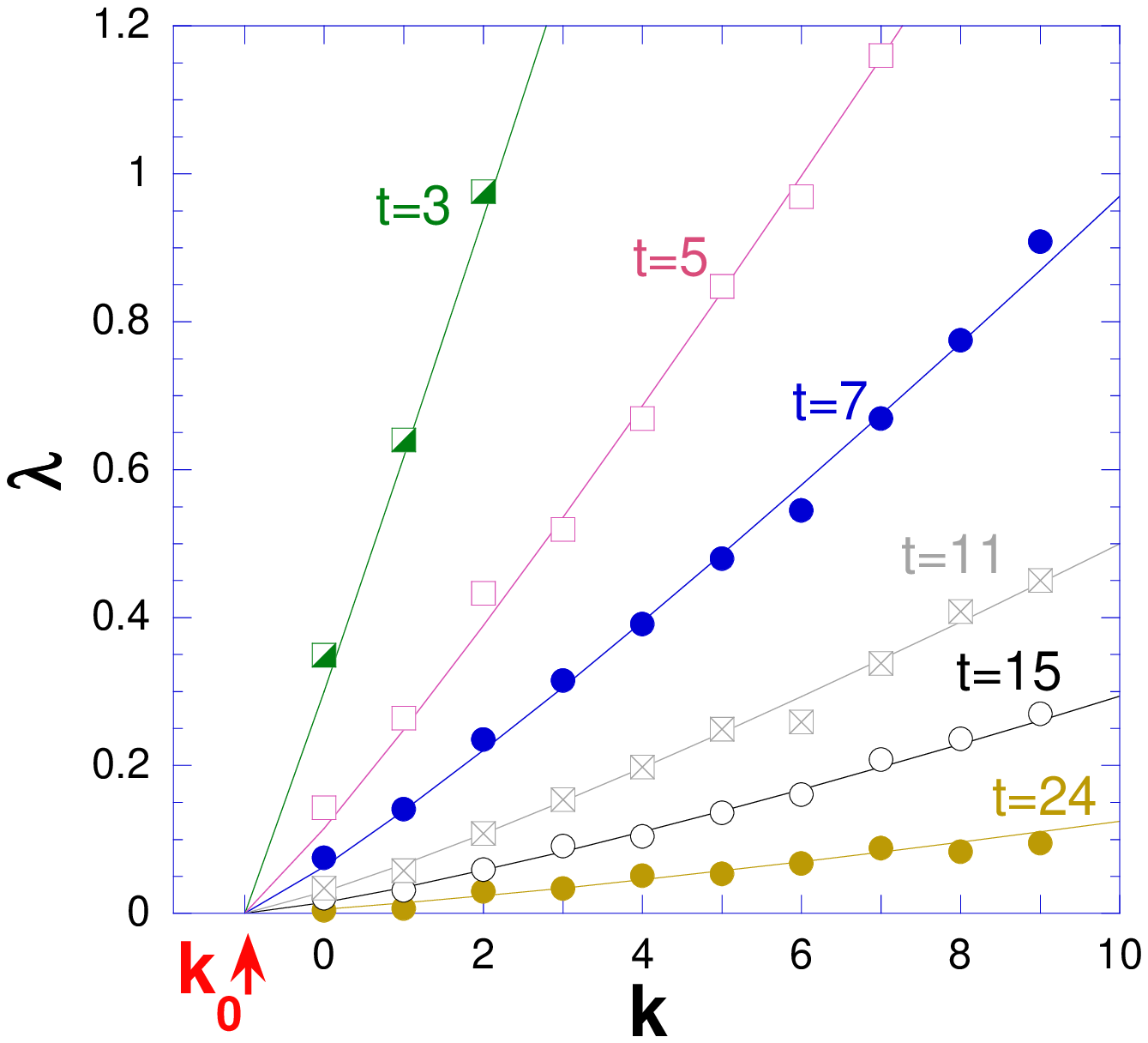}

\caption{Left panel: mean annual citation rate, $\lambda(k)=\overline{\Delta k_{i}}$, as a function of the  number of previous citations  $k$; $t$ is the number of years after publication.   The continuous lines show superlinear fit, $\lambda=A(k+k_{0})^{\alpha}$ where $k_{0}=1$ and  $\alpha$ is shown in the inset. Right panel: The same data in the linear scale. The intercept of the continuous lines with the  horizontal axis yields time-independent $k_{0}\approx 1$.
}
\label{fig:mean}
\end{figure}

Figure \ref{fig:mean} shows that $\lambda(k)$ dependence is well accounted for by  Eq.\ref{rate} where $A,k_{0}$, and $\alpha$ are fitting parameters. We found that the aging function follows the power-law decay, $A=3.54/(t+0.3)^2$; the initial attractivity is almost time-independent, $k_{0}\approx 1.1$; (iii) the exponent $\alpha$ gradually increases with time from $\alpha=1$ to  $\alpha=1.25$. Although the deviation of  $\alpha$ from unity is small, it is significant and contrasts the  assumption of linearity commonly accepted  by the practitioners of the preferential attachment model \cite{Barabasi,Simon,Solla,Medo,Valverde}. Indeed, while the linear preferential attachment generates the scale-free network with the power-law degree distribution,  the superlinear preferential attachment tends to generate  the "winner-takes-all"  network \cite{KRL,Dorogovtsev2002}.

For comparison, we performed similar measurements for the Mathematics and Economics papers  published in the same year (1984). We found that the citation dynamics for both these disciplines is also well accounted for by Eq.\ref{rate}. The $\alpha$ and $k_{0}$ turn out to be almost the same as those for Physics while the aging function $A(t)$ is different (see Supplementary Material). Similar $\alpha$ and $k_{0}$ were found in the US patent citation studies \cite{Csardi}. This suggests universal microscopic  mechanism of citation accumulation whereas the variations in total citation counts between scientific fields can be attributed to different initial conditions  (the number of citations gained during first couple of years after publication) and to different growth rates of the number of publications.

In what follows we analyze  another key ingredient of the  preferential attachment model - the Markov chain assumption. Since Eq.\ref{rate} postulates that the  citation rate $\lambda=\overline{\Delta k_{i}}$ depends \textit{only} on the number of previous citations $k$, it follows that the statistical distribution of additional citations $\Delta k_{i}$, gained by the papers with the same $k$ during a time window $\Delta t$,  should be nothing else but the Poissonian: 
\begin{equation}
P(\Delta k)=e^{-\lambda \Delta t}\frac{(\lambda \Delta t)^{\Delta k}}{(\Delta k)!}.
\label{Poisson}
\end{equation}

\begin{figure}
\includegraphics[width=0.45\textwidth]{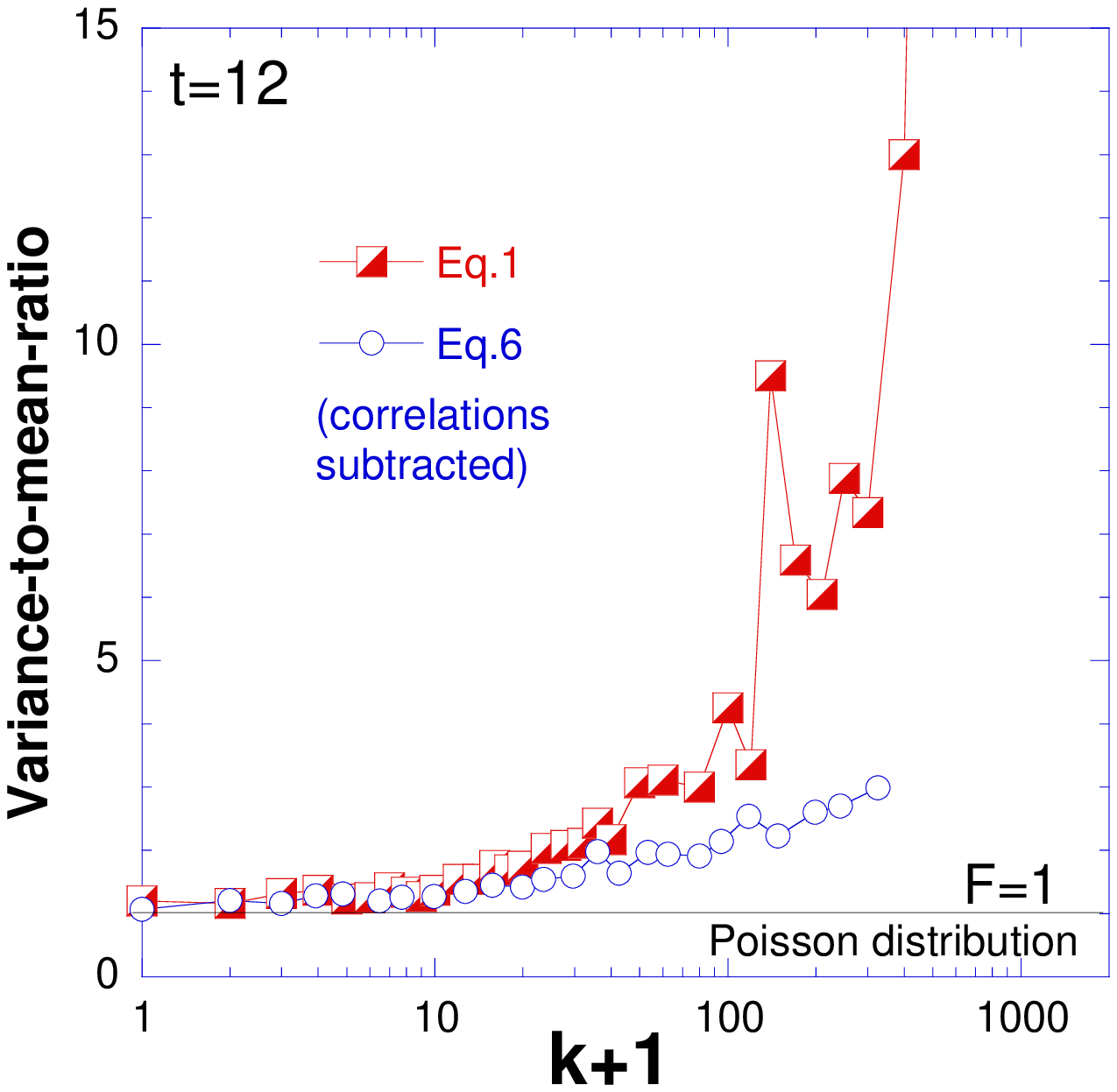}%0.7
\includegraphics[width=0.45\textwidth]{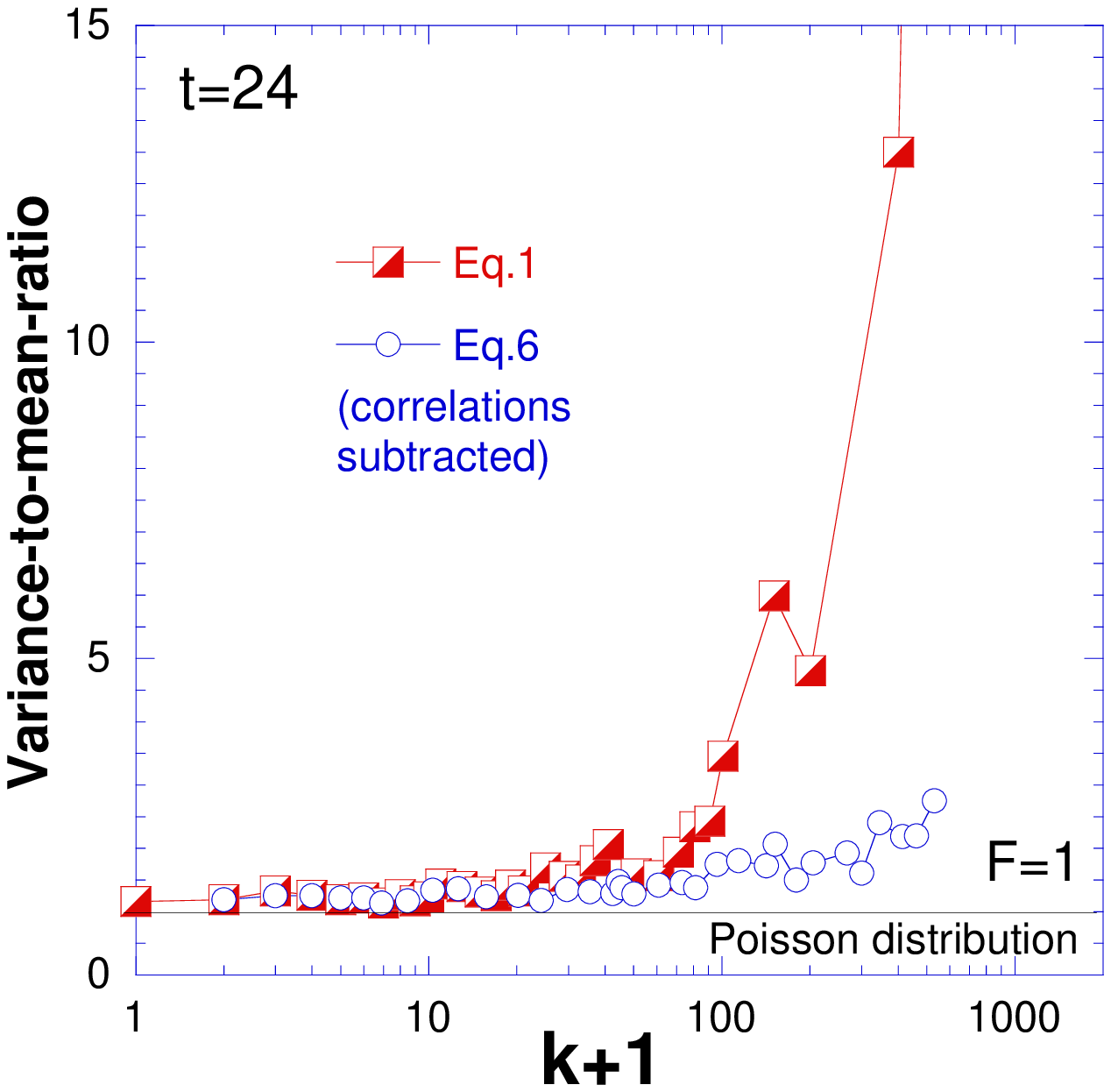}%0.7
\caption{The variance-to-mean ratio (Fano factor), $F=\sigma^{2}/\lambda$,  for the statistical distributions of additional citations $\Delta k_{i}$ (see Fig.\ref{fig:NG}). Each point corresponds to the set of papers with the same  number of expected citations $\lambda_{i}$, given by Eq.\ref{rate} (the red squares). The data, especially for $k>60$, deviate upwards from the $F=1$ line, characteristic for the Poisson distribution.  The  blue circles show the variance-to-mean ratio for the $\Delta k_{i}$-distributions for the sets of papers with the same number of expected citations $\lambda_{i}$, Eq.\ref{rate2}. These data are closer to the $F=1$ line. 
}  
\label{fig:Fano}
\end{figure}

\begin{figure}
\includegraphics[width=0.5\textwidth]{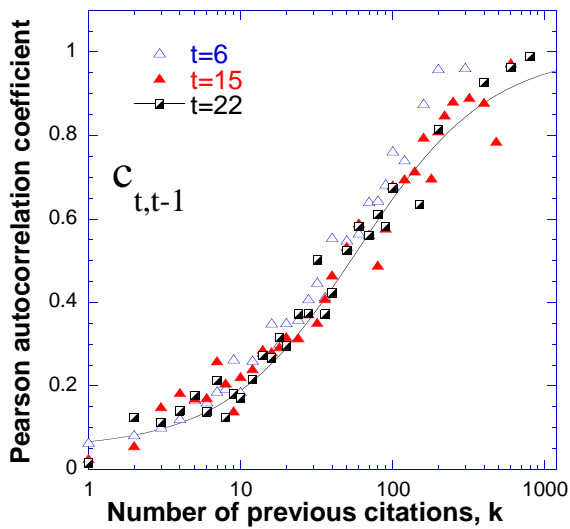}
\caption{ The Pearson autocorrelation coefficient  for additional citations (Eq.\ref{c}). Each point corresponds to the set of papers with the same  number of previous citations  $k$ garnered by a certain citing year $t$. The data for different $t$ almost collapse. The continuous line shows empirical approximation (Eq.\ref{c-approx}). 
 }
\label{fig:Pearson}
\end{figure}

To the best of our knowledge, statistical distribution of additional citations was not measured so far. This new  kind of measurements  (Fig.\ref{fig:NG}) reveals that  the $\Delta k_{i}$-distributions  are broader than the Poissonian. To quantify  this broadening  we used the variance-to-mean ratio, $F=\sigma^{2}/\lambda$,  also known as index of dispersion or Fano factor.  Figure \ref{fig:Fano}  shows that  $F\approx 1$ for small $k$, as expected for the Poisson distribution, while  $F>>1$ for large $k$. 
This strong deviation from the Poissonian indicates that  Eq.\ref{rate} misses some important factor which determines the growth of  citation networks. We reasoned that the missing factor is  related to  citation history of papers. To probe this conjecture we considered the temporal autocorrelation of the annual citations, $\Delta k_{i}(t)$. Since the typical citation history of a paper is too short (10-15 years), the measurement of autocorrelation for a single paper is unreliable. Therefore, we measured autocorrelation in the sets of papers that at certain citing year $t$ have  the same number of previous citations $k$. Specifically, we found  the number of citations garnered by each paper in such set during the current year and the last year-  $\Delta k_{i,t}$ and $\Delta k_{i,t-1}$, correspondingly, and calculated the Pearson autocorrelation coefficient  
\begin{equation}
c_{t,t-1}=\frac{\overline{\left(\Delta k_{i,t}-\overline{\Delta k_{i,t}}\right)\left(\Delta k_{i,t-1}-\overline{\Delta k_{i,t-1}}\right)}}{\sigma_{t}\sigma_{t-1}}
\label{c}
\end{equation}
Here,  $\sigma_{t},\sigma_{t-1}$ are the standard deviations of the $\Delta k_{i,t}$ and $\Delta k_{i,t-1}$ distributions, respectively  ($\sigma_{t}\approx\sigma_{t-1}$), and the averaging is performed over all papers in the set. This was done for all $k$ and $t$. Figure \ref{fig:Pearson} shows that $c_{t,t-1}$ grows with $k$.  For  moderately cited papers, $k<<60$, the autocorrelation is weak while for  highly cited papers, $k>>60$, the autocorrelation is strong: $c\sim 1$. 
The empirical function
\begin{equation}
c(k,t)\approx\frac{k+3}{k+60}
\label{c-approx}
\end{equation}
fits well our measurements. Strong temporal autocorrelation of citations  violates the underlying assumption of the  preferential attachment  model \cite{Barabasi,Solla,Simon}:  it turns out that citations dynamics is not a Markov process since it depends on past history.

We suggest a more realistic growth model that is based on the the first-order linear autoregression,
 \begin{equation}
\lambda_{i}= (1-c)A(k_{i}+k_{0})^{\alpha}+c\Delta k_{i,t-1}
\label{rate2}
\end{equation}
where  $\lambda$ is the latent citation rate and $c$ is given by Eq.\ref{c-approx}. The actual number of citations is given by Eq.\ref{Poisson}. Equation \ref{rate2} introduces  positive feedback between successive citations of the same paper, in other words, it approximates  the citation dynamics of a paper by the inhomogeneous self-exciting point process \cite{Hawkes}.  (Similar ideas were discussed in Refs. \cite{Wang2008,Cattuto}.) The resulting preferential attachment model replaces Eq.\ref{rate} by Eq.\ref{rate2} in such a way that the  stochastic term in Eq.\ref{delta-k} reduces to the Poissonian noise.   Equation \ref{rate2} states that the latent citation rate of a paper \cite{Burrell}  depends not only on the total number of accumulated citations but on the recent citation rate as well. This accounts for the  "sleeping beauties": the papers that initially  had  small number of citations but suddenly became popular. While the conventional preferential attachment model (Eq.\ref{rate}) yields predominance of the  "first-movers" \cite{mover}, our more realistic model allocates a fair share of citations to "sleeping beauties".

%The "sleeping beauties"   compete with  "first-movers" \cite{mover} that, according to the conventional preferential attachment model (Eq.\ref{rate}), should capture the whole dynamics of citation network.  %Of course, for the papers whose recent citation rate is close to average, $\Delta k_{i,t-1}=A(k_{i}+k_{0})^{\alpha}$, Eq. \ref{rate2} reduces to Eq.\ref{rate}.

\begin{figure}
\includegraphics[width=0.45\textwidth]{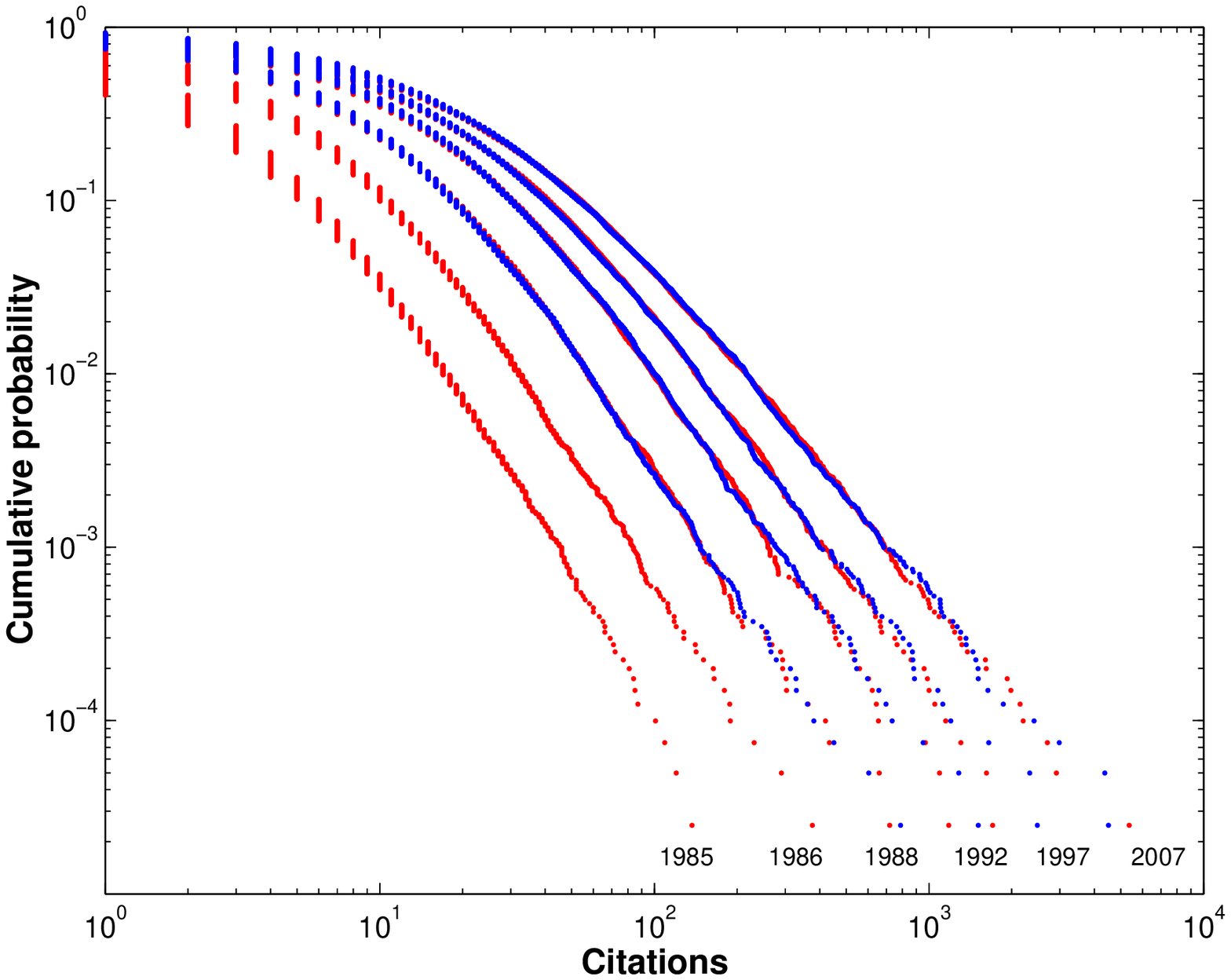}
\includegraphics[width=0.45\textwidth]{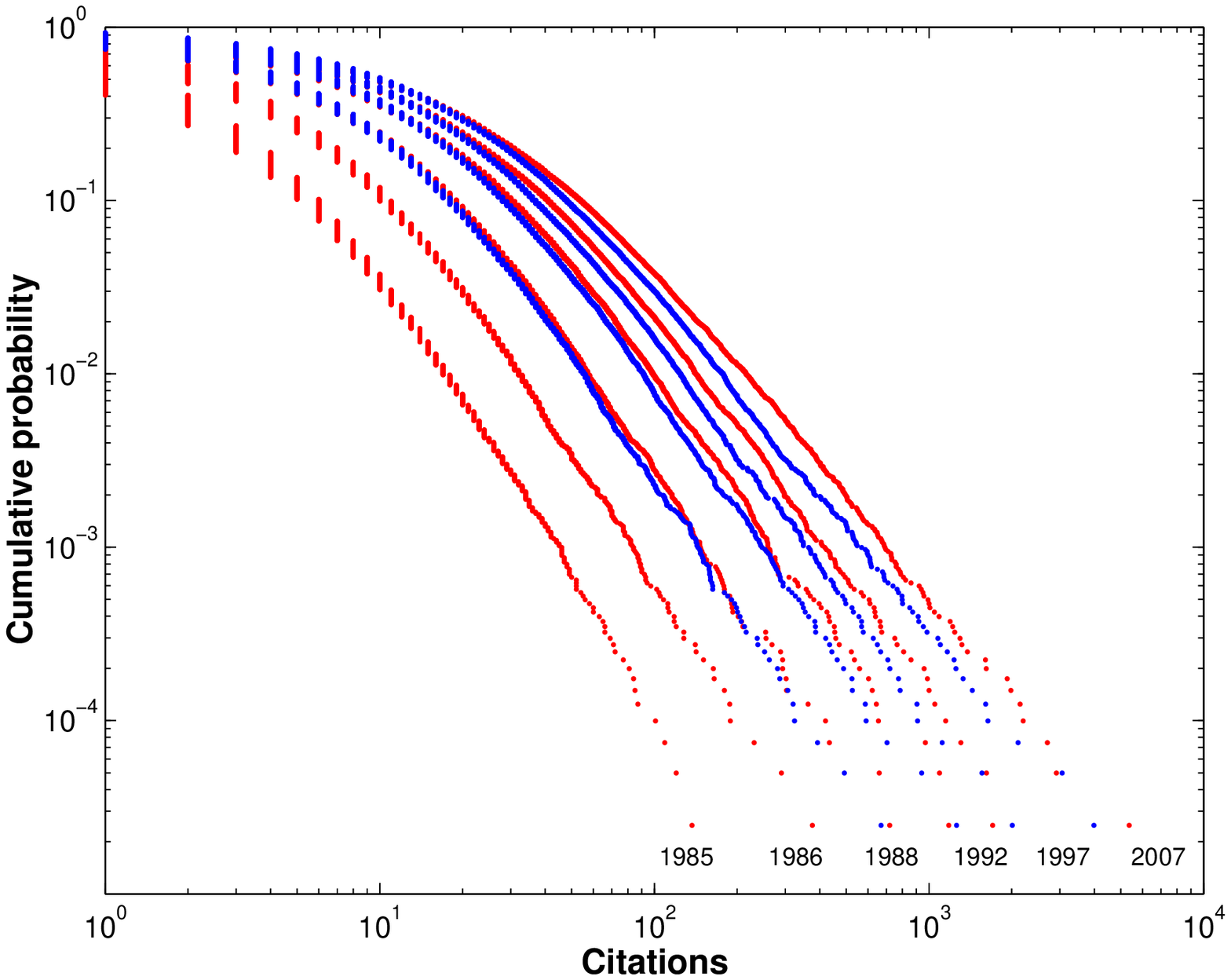}
\caption{Cumulative citation distributions for 40,195 Physics papers published in 1984. The citing year is indicated at each curve. Red symbols - measurements, blue symbols- numerical simulation assuming initial citation distributions  of 1985 and 1986. Left panel: Full model based on Eq.\ref{Poisson} and Eq.\ref{rate2} provides  excellent fits to the measured citation distributions. Right panel: Incomplete model  based on Eq.\ref{rate}(correlations ignored) and Eq.\ref{Poisson} underestimates citation counts.
%Mean number of additional citations, $\overline{\Delta k_{i}}$, versus prediction of Eq.\ref{rate2}. The straight line  shows $\lambda_{i}=\overline{\Delta k_{i}}$ dependence. 
}
\label{fig:simulation}
\end{figure}

To verify the multiplicative stochastic model described by Eq.\ref{rate2} we chose all Physics papers published in the same year (1984), fixed a certain citing year ($t=$1986),  measured the number of total and last year citations, $k_{i,t}$ and $\Delta k_{i,t-1}$,  and calculated $\lambda_{i}$ for each paper using Eq.\ref{rate2} with experimentally measured parameters $c(k),A(t),k_{0},$ and $\alpha(t)$. Then we run numerical simulations assuming Poisson process with the rate given by Eq.\ref{rate2}, found the number of citations of each paper in the year $t+1$, and calculated  the  cumulative distribution of citations. The procedure was repeated for the next year and so on. Figure \ref{fig:simulation}a shows that this algorithm  closely reproduces the actual citation distribution for each citing year. This means that Eq.\ref{rate2} yields an excellent description not only of the microscopic citation dynamics but of the macroscopic citation distribution as well.  On another hand, the numerical simulation
 that assumes only Poisson process and ignores correlations, does not reproduce well our measurements  (Fig. \ref{fig:simulation}b).  

What are the implications of our study? We find  that the cumulative citation distribution is neither stable nor stationary but develops in time. Immediately after publication the  spread of initial conditions (journal circulation numbers) yields convex cumulative distribution of citations that can be fitted equally well by the (discrete) power-law \cite{Redner1998,Wallace,Peterson} or log-normal \cite{Redner2005,Amaral,Fortunato,Petersen} functions. Thereafter, citation dynamics of most papers is dominated by the first term in Eq.\ref{rate2} in such a way that the citation history of papers that managed to garner less than 50-70 citations is completed after 10-15 years.  However, the papers with more than 50-70 citations  continue to be cited even after 10-15 years, their dynamics being determined by the second term in Eq.\ref{rate2} which does not decay with time. In other words, while the bulk of the citation distribution becomes stable,   the tail grows. In the course of time its shape changes from the convex to concave  in such a way that for the most part of the time the tail looks straight in the log-log coordinates. Although such power-law tail was previously considered as a fingerprint of the scale-free network,  at least for citation network it turns out to be a transient phenomenon. The intrinsic scale of the citation network, $k_{cr}=50-70$, is clearly revealed  in the microscopic dynamics (Fig.\ref{fig:Pearson}). We conclude that the almost power-law degree distribution of  citations  that was previously interpreted as the indication of the scale-free network  \cite{Barabasi,Newman_SIAM,Solla,Redner1998} arises from the interplay between   aging \cite{Dorogovtsev-aging}, multiplicative stochastic process (Eq.\ref{delta-k}), and superlinear preferential attachment.
% that tends to generate  the "winner-takes-all" citation network \cite{KRL,Dorogovtsev2002}. An indication on the latter  was indeed found in our recent measurements \cite{Golosovsky}.

The two-term Eq. \ref{rate2} implies that scientific papers constitute two broad classes with respect to their longevity \cite{Redner2005}. The citation rate of the 90$\%$ of the  papers  achieves its maximum  in 2-3 years after publication and  decays to zero in 10-15 years. Citation dynamics of these papers is the aftereffect of their initial hit and is more or less predictable since the impact of these papers is probably limited to several research groups and does not propagate further.  However, citation rate of $10\%$ of the papers that overcome the tipping point \cite{Peterson} of $k_{cr}\simeq 50-70$ citations   is determined more by their recent citation history. It seems that these papers have a continuing impact \cite{Golosovsky}which propagates from one research group to another in a cascade process like in epidemics \cite{Goffman}. This  diffusion of scientific knowledge \cite{Jaffe} extends the paper longevity to  much more  than 10-15 years.

In summary, our measurements indicate that the  mechanism that generates complex networks may be more sophisticated than the memoryless linear preferential attachment assumed so far. We propose a  stochastic growth model  that considers the evolution of the node degree as an inhomogeneous self-exciting point process. In the context of citations, the model is fully verified by our microscopic and macroscopic  measurements  and  can serve for  prognostication of the future citation behavior of a paper, group of papers, or of a journal impact factor.

We are grateful to S. Redner, N. Shnerb, and A. Scharnhorst  for insightful discussions.

\section*{Supplementary material: Methodology}
To find time evolution  of the number of citations of Physics papers we used the Thomson-Reuters ISI Web of Science. We chose a certain publishing year (1984) that on the one hand  is distant enough from now, while on another hand, the contents of the most part of the papers published in this year is available now in the electronic format. We considered the fields of Physics, Astrophysics, Optics, Crystallography, and Material Science  and excluded popular science and review journals.  We performed the search using 82 journals with the highest annual number of publications $N$ and  considered only articles, letters and notes while the  editorial material, comments, and reviews were excluded. To find Physics papers in the multidisciplinary Nature and Science journals we looked through the titles of all publications in 1984 and chose only those  that by our opinion fall under Physics category. 

\begin{figure}[ht]
\includegraphics[width=0.25\textwidth]{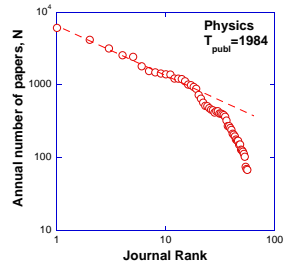}
\caption{The rank-size plot of the Physics journals where the rank is the  number of papers published in a each journal in 1984. The dashed line shows the power-law fit  representing the Bradford's law.}
\label{fig:Bradford}
\end{figure}

How representative is this list of journals?  To answer this question we invoke Bradford's law \cite{Bradford} stating that few largest journals  contain the dominant part of papers in the field. Figure \ref{fig:Bradford} shows that the rank-size distribution for the set of  largest 82 journals  exhibits the  power-law dependence with cut-off. Extrapolation to  $N=1$ yields that these  82 journals contain $\sim 95 \%$ of all Physics papers published in 1984. 

\section*{Supplementary material: The list of Physics journals}
Physical Review A,B,C,D; Physical Review Letters; Japanese Journal of Applied Physics A,B; Journal of Physics*; Physics Letters A,B; Acta Crystallographica*; Journal of Chemical Physics; Physica Status Solidi*; Nuclear Instruments*; Astrophysical Journal; Journal of Physical Chemistry; Journal of Applied Physics; Journal de Physique*; Chemical Physics Letters; Nuclear Physics*; Journal of the Optical Society of America*; Solid State Communications; Zeitschrift fur Physik*; Applied Physics Letters; Applied Optics; Surface Science; Journal of the Physical Society of Japan; Journal of  Vacuum Science*; Journal of Non-Crystalline Solids; Journal of Crystal Growth; Comptes Rendu de'l Academie des Sciences Serie II *; Thin Solid Films; Journal of Materials Science; American Journal of Physics; Physics of Fluids; Nuovo Cimento*; Review of Scientific Instruments; Physica ?; Chemical Physics; Optics Communications; Journal of Magnetism and Magnetic Materials; Journal of Luminescence; Journal of Fluid Mechanics; Philosophical Magazine*; Physica Scripta; Canadian Journal of Physics; Journal of Statistical Physics; Optics Letters; Applied Spectroscopy*; Solid State Ionics; Plasma Physics and Controlled Fusion; Journal of Physics and Chemistry of Solids; Journal of Computational Physics; Communications in Mathematical Physics; Zeitschrift fur Kristallographie*;
Nature, Science. 
%%%%%%%%%%%%%%%%%%%%%%%%%%%%%%%%%%%%%%%%%%%%%%%%%%%%%%%%%%%%%%%%

\section*{Supplementary material: Homogeneity of Physics}

The homogeneity of our data set with respect to citations is ensured by the fact that the papers in different Physics subdisciplines are published in the same journals. These journals  impose a certain standard of the  reference list length that provides a natural scale for  citations in this scientific  discipline. Indeed, assume the cohort of all papers  that were published in the same year and consider citation distribution for this cohort after many years. Since all these papers will be cited predominantly by the papers in the same discipline, and neglecting the $2\%$ annual growth of the number of publications, the mean number of citations would be equal to the average length of the reference list. This condition of stationarity was the rationale for our choice of the whole Physics discipline rather than a single subdiscipline.

\begin{figure}[h]
\includegraphics[width=0.23\textwidth]{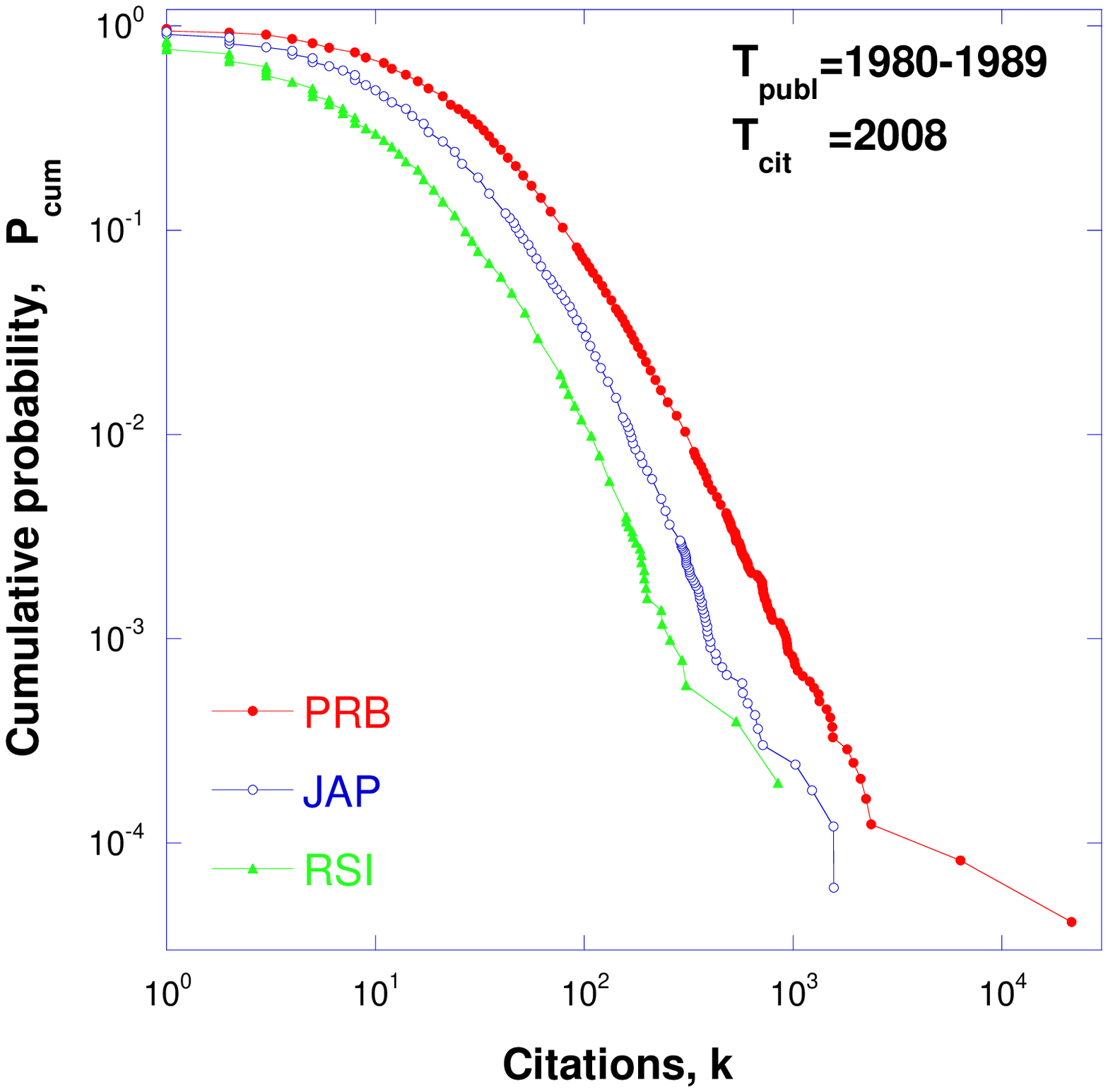}
\includegraphics[width=0.23\textwidth]{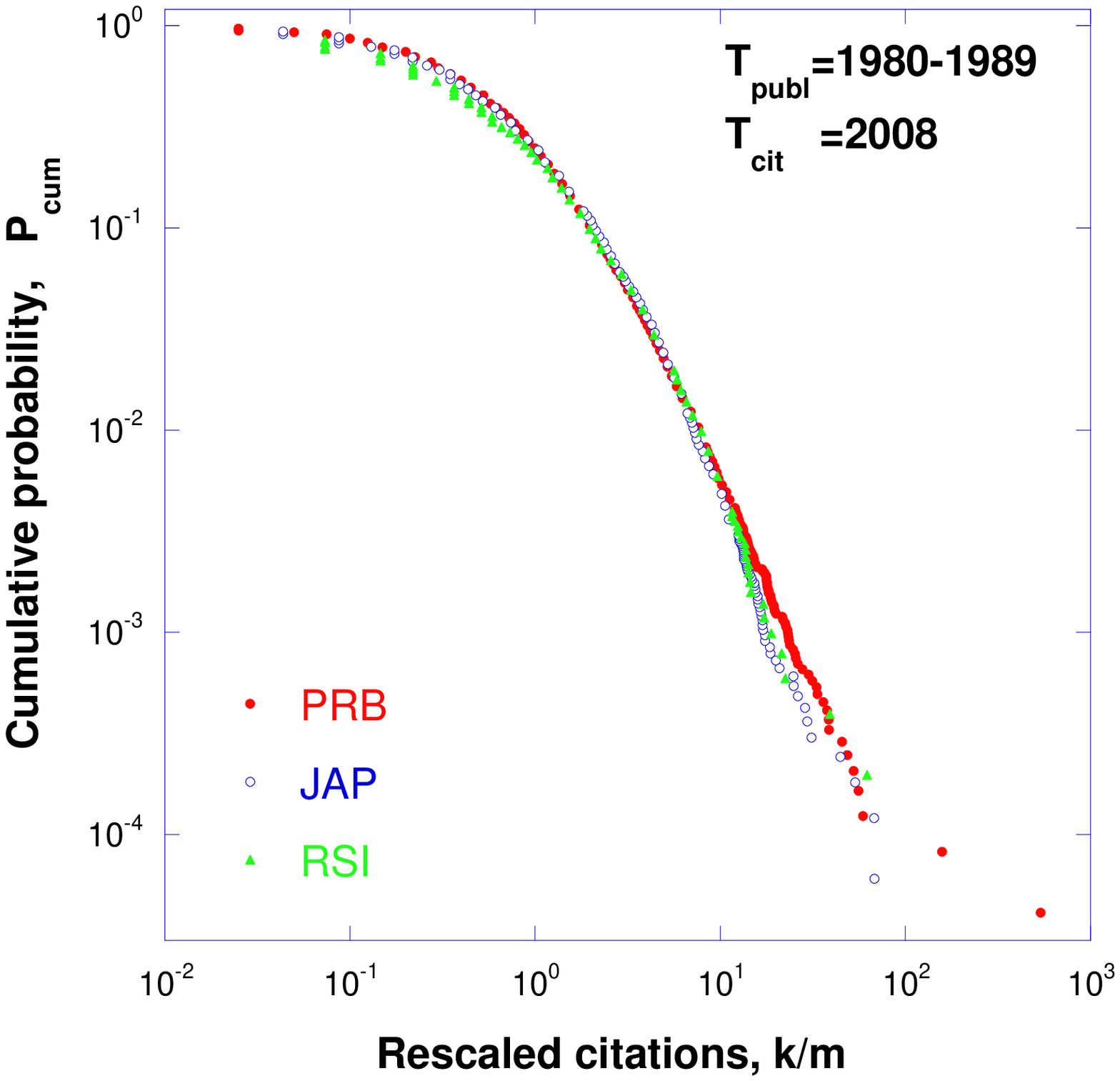}
\caption{Cumulative distribution of citations to the papers published in 1980-1989 and cited by 2008.  We show results for the Physical Review B (PRB), Journal of Applied Physics (JAP) and Review of Scientific Instruments  (RSI). Although citation distributions for these three large journals look different (left panel), the difference is in  scale and not in  shape. The scaled distributions (right panel) are much more alike.
}
\label{fig:journals}
\end{figure}

This homogeneity of Physics with respect to citations is  also revealed from the inspection of the cumulative citation distributions for the papers published in different Physics journals. %\subsection{Comparison between different Physics journals} The short answer is as follows: to predict citation dynamics of a paper we need dynamic equation (growth model) and initial conditions. The dynamic equation  appears to be universal while the initial conditions (the number of citations garnered during first couple of years after publication and the initial citation rate) can differ from one subdiscipline to another. 
Indeed, in the researcher's perception the theoretical and experimental papers could have different citation pattern. To find out to what extent this is true we chose three  large Physics journals  that do not have page limitation- Physical Review B (PRB), Journal of Aplied Physics (JAP), and Review of Scientific Instruments (RSI). As seen from the reader/author's perspective,  PRB publishes theoretical and experimental papers in fair proportion, JAP is more biased towards experimental papers, and RSI publishes almost exclusively experimental papers. Figure \ref{fig:journals} shows cumulative citation distribution for the papers published in these journals in the decade 1980-1989. Although these distributions look different they almost collapse after rescaling $k\rightarrow k/m$ where $m$ is a journal-specific constant \cite{Fortunato}. This means that the growth model for citations  of the papers published in these three journals is actually the same and only the scale (initial conditions, namely, the number of citations garnered by a paper during first 2-3 years after publication) is different. 

The homogeneity of the Physics discipline with respect to citations becomes evident also from the consideration of the list of the most cited Physics papers. As an example we chose fifteen most cited papers published in the decade of 1980-1989 and cited by 2008. Notably, the list below is  not dominated by a single subdiscipline. Although the theoretical papers prevail, the experimental papers are represented fairly well (26.6$\%$). (The absence of  several Nobel prize-winning papers that appeared during this period  is explained by the fact that they quickly became common knowledge). 
  
 \begin{itemize}
\item  C.T. Lee, W.T. Yang, R.G. Parr, ''Development of the Cole-Salvetti \MakeLowercase{CORRELATION-ENERGY FORMULA INTO A FUNCTIONAL OF THE ELECTRON-DENSITY}'', \emph{Phys.Rev.B}   (1988);  22031 citations.   
\item A.D. Becke, ''D\MakeLowercase{ENSITY-FUNCTIONAL EXCHANGE-ENERGY APPROXIMATION WITH CORRECT ASYMPTOTIC- BEHAVIOR}'',  \emph{Phys.Rev.A }(1988);  14383 cit.
\item  N. Walker, D.Stuart, ''A\MakeLowercase{N EMPIRICAL-METHOD FOR CORRECTING DIFFRACTOMETER DATA FOR ABSORPTION EFFECTS}'',  \emph{Acta Crystallographica A}  (1983);  8499 cit.    
\item  S.H. Vosko, L. Wilk, M. Nusair, ''A\MakeLowercase{CCURATE SPIN-DEPENDENT ELECTRON LIQUID CORRELATION ENERGIES FOR LOCAL SPIN-DENSITY CALCULATIONS - A CRITICAL ANALYSIS}'', \emph{Canadian Journal of Physics} (1980);  8234 cit.                                     
\item T.H. Dunnig,  ''G\MakeLowercase{AUSSIAN-BASIS SETS FOR USE IN CORRELATED MOLECULAR CALCULATIONS .1. THE ATOMS BORON THROUGH NEON AND HYDROGEN}'', \emph{J. Chem.Phys.} (1989); 8186 cit.    
\item  J.G. Bednorz, K.A. Muller, ''Possible high-$T_{C}$ superconductivity in the Ba-La-Cu-O system'',  \emph{Zeitschrift fur Physik-Condensed matter}, (1986); 7407 cit.  
 \item  S.Kirkpatrick, C.D. Gelatt, M.P. Vecchi,  ''O\MakeLowercase{PTIMIZATION BY SIMULATED ANNEALING}'', \emph{Science} (1983); 6907 cit.  
\item W.L. Jorgensen, J. Chandrasekhar, J.D. Madura, R.W. Impey, M.L. Klein, ''C\MakeLowercase{OMPARISON OF SIMPLE POTENTIAL FUNCTIONS FOR SIMULATING LIQUID WATER}'',   \emph{J.Chem.Phys.} (1983); 6838 cit.    
\item J.P.Perdew, A. Zunger, ''S\MakeLowercase{ELF-INTERACTION CORRECTION TO DENSITY- FUNCTIONAL APPROXIMATIONS FOR MANY-ELECTRON SYSTEMS}'', \emph{ Phys.Rev. B} (1981); 6784 cit.    
\item J.P.Perdew, ''D\MakeLowercase{ENSITY-FUNCTIONAL APPROXIMATION FOR THE CORRELATION-ENERGY OF THE INHOMOGENEOUS ELECTRON-GAS}'',   \emph{Phys.Rev.B }(1986); 6429 cit.   
\item H.D.Flack,  ''O\MakeLowercase{N ENANTIOMORPH-POLARITY ESTIMATION}'', \emph{Acta Crystallographica A } (1983);  6381 cit.    
\item H.J.C. Berendsen, J.P.M. Postma, W.F. Vangusteren, A.Dinola, J.R. Haak, ''M\MakeLowercase{OLECULAR-DYNAMICS WITH COUPLING TO AN EXTERNAL BATH}'', \emph{J. Chem.Phys.} (1984);  5877 cit. 
\item H.W. Kroto, J.R. Heath, S.C. Obrien, R.F. Curl, R.E. Smalley, ''C-60\MakeLowercase{ - BUCKMINSTERFULLERENE}'',  \emph{Nature} (1985);  5861 cit. 
\item  D.M. Ceperley, B.J. Alder,  ''G\MakeLowercase{ROUND-STATE OF THE ELECTRON-GAS BY A STOCHASTIC METHOD}'', \emph{Phys.Rev.Lett. }(1980); 5499 cit. 
\item  G. Binnig, C.F. Quate, C. Gerber, ''A\MakeLowercase{TOMIC FORCE MICROSCOPE}'', \emph{Phys.Rev.Lett.} (1986); 5364 cit.  
\end{itemize}
\section*{Supplementary material: Comparison to other  disciplines}
We performed similar studies using Mathematical and Economics papers. Unlike  Physics study where we tried to cover the whole field in order to achieve good statistics, here we adopted a different choice and considered only pure Mathematical and pure Economics papers. The resulting data set is more homogeneous but does not have enough  statistics due to relatively small number of papers. We used  the Thomson Reuters ISI Web of Science, chose all relevant  Math journals (500 titles), considered all original research Math papers published in these journals in one year -1984 excluding review articles, comments, editorial, etc.  This left  6313 Math papers. A similar search in the field of Economics (165 relevant journals) yielded 3043 original research papers  published in 1984. We analyzed citation history of these papers from 1984 to 2011.

\begin{figure}[h]
\includegraphics[width=0.23\textwidth]{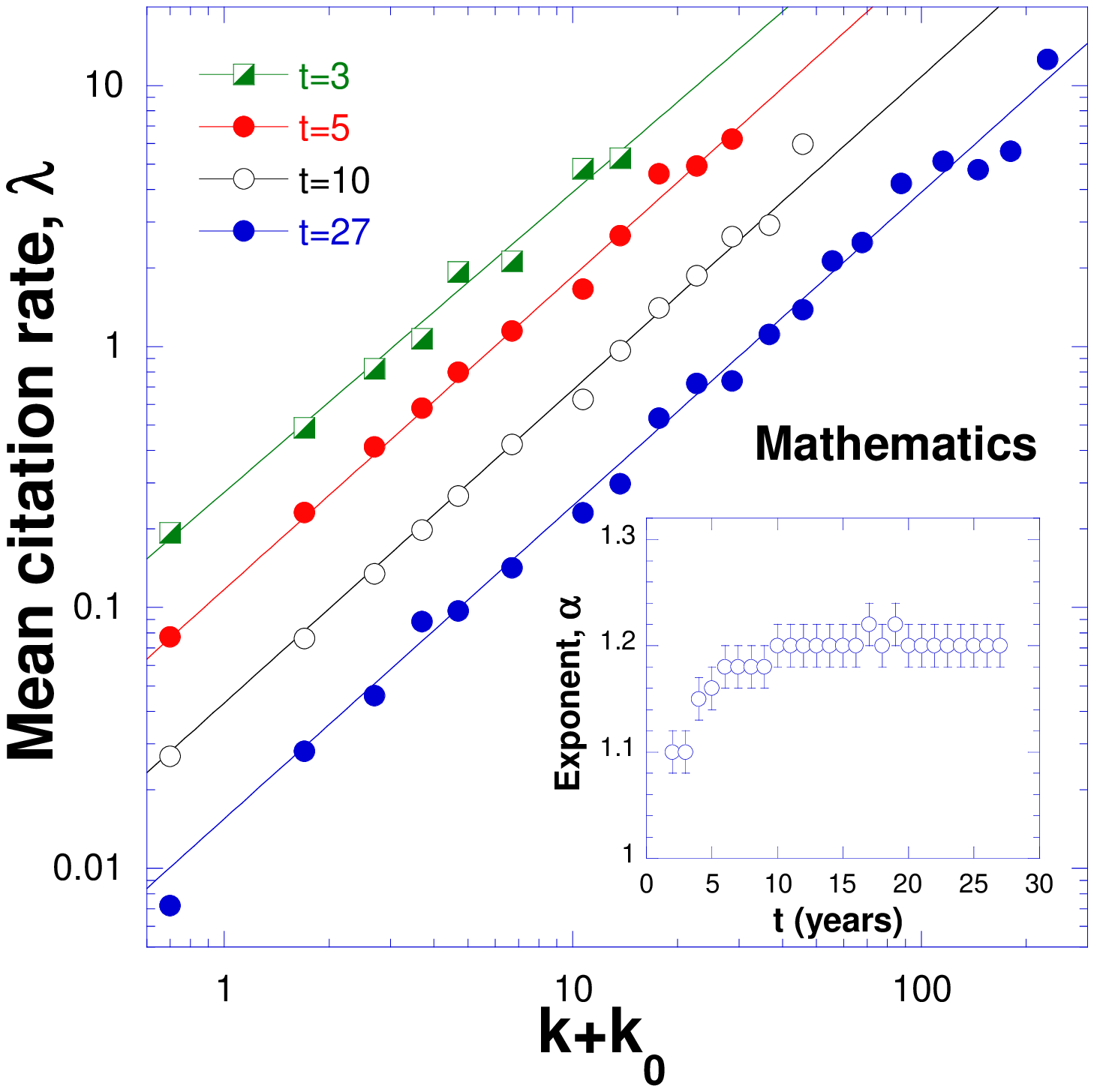}
\includegraphics[width=0.26\textwidth]{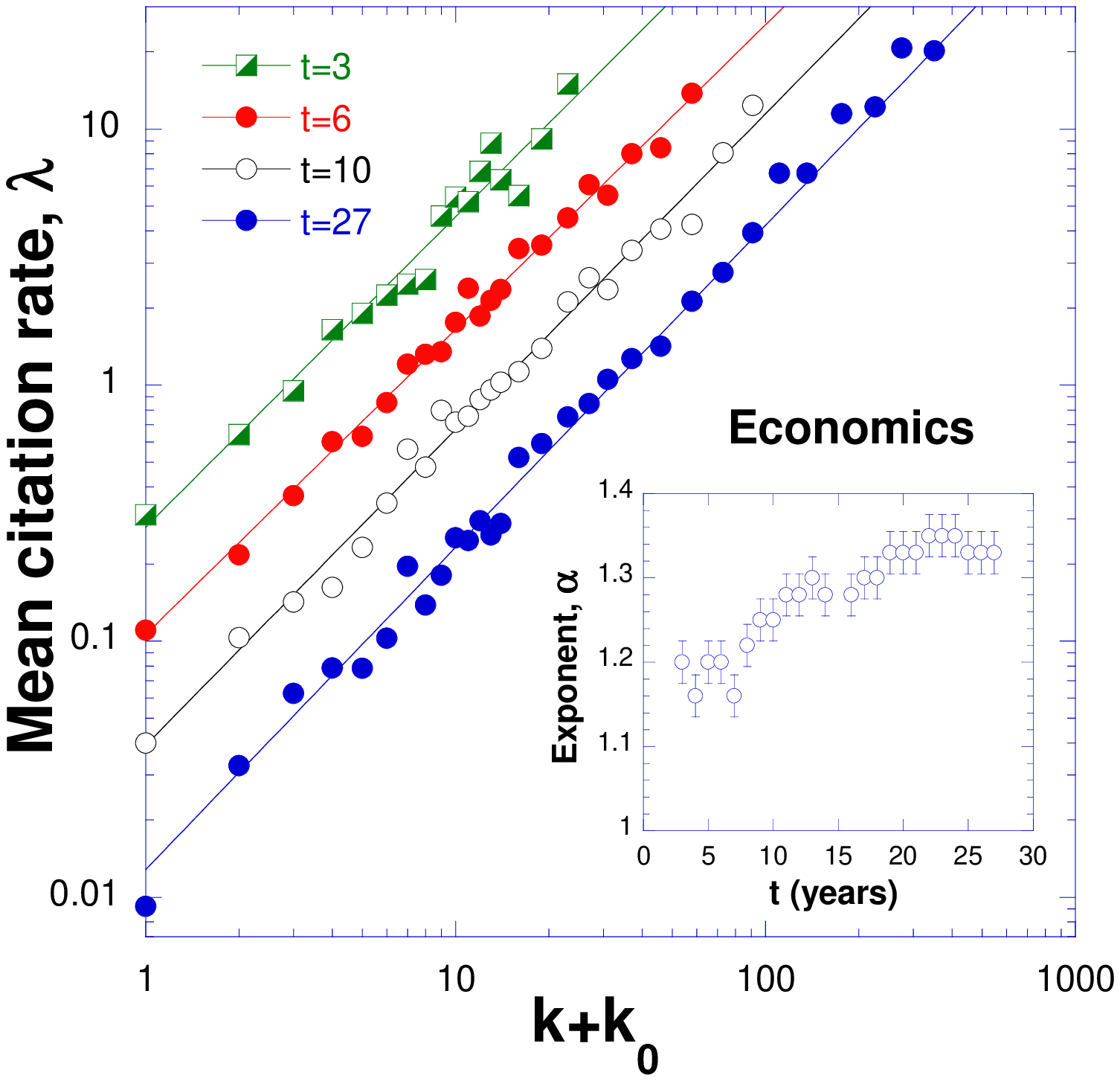}
\includegraphics[width=0.25\textwidth]{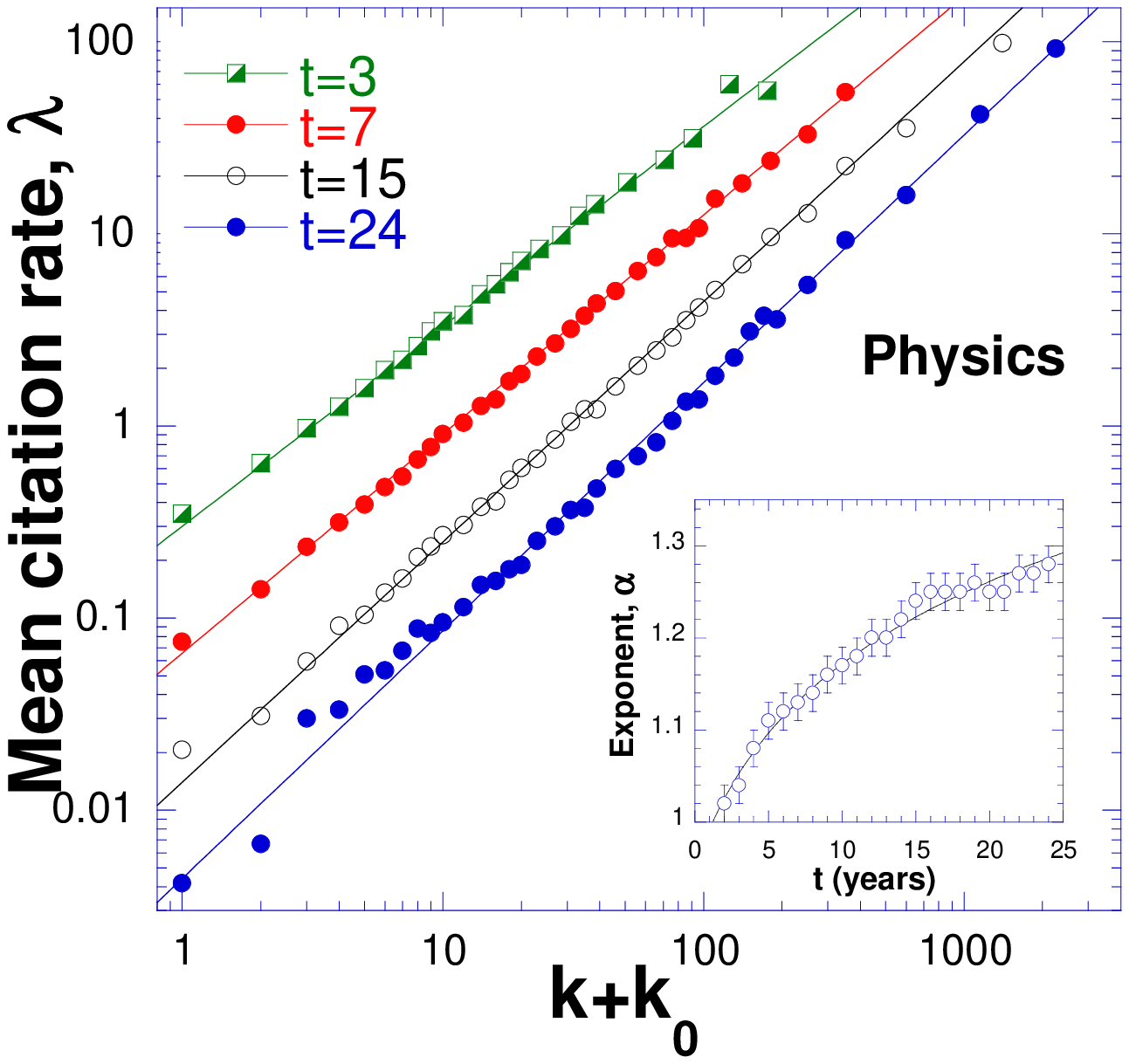}
\caption{Mean annual citation rate, $\lambda(k)=\overline{\Delta k_{i}}$, as a function of the  number of previous citations  $k$ for  the Mathematical,  Economics, and Physics  papers - all published in 1984. $t$ is the number of years after publication.   The continuous lines show superlinear approximations, $\lambda=A(k+k_{0})^{\alpha}$ where $k_{0}=0.7$ for Mathematics, $k_{0}=1$ for Economics, and $k_{0}=1.1$ for Physics. The exponent  $\alpha(t)$ is time-dependent as shown in the insets. Note different  scales  in (a),(b), and (c).
}
\label{fig:lambda}
\end{figure}
For each paper $i$  we determined $k_{i,t}$ - the total number of  citations accumulated  after  $t$ years ($t=T_{cit}-T_{publ}+1$),  and $\Delta k_{i,t}$- the number of additional citations gained by the paper in the  year  $t+1$.  We  chose a certain citing year $t$ and grouped all papers into  logarithmically-spaced bins, each bin containing the papers with close $k$, and found the  mean, $\lambda(k)=\overline{\Delta k_{i}}$, of each  distribution. 

Figure \ref{fig:lambda} shows that the $\lambda(k)$ plots for all three disciplines can be  fitted by the superlinear dependence 
\begin{equation}
\lambda=A(k+k_{0})^{\alpha}
%\Delta k_{i}=\lambda(k,t)+\epsilon
\label{rate}
\end{equation}
where $k$ is the number of previous citations,   $k_{0}$ is the  "initial attractivity",  $\alpha$ is the attachment exponent, and $A(t)$ is the aging function.  
\begin{figure}[h]
\includegraphics[width=0.3\textwidth]{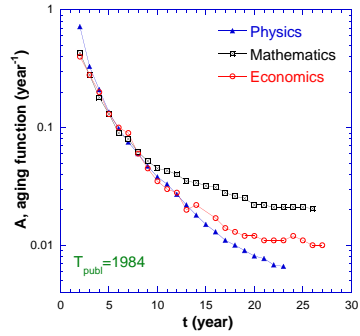}
\caption{$A$, the aging function  for the Mathematical, Economics and Physics papers. The data for the first couple of years are different, the data for subsequent 6-8 years are close and  10 years after publication the data diverge once again. 
}
\label{fig:A}
\end{figure}

We observe that the citation dynamics for all three disciplines follows superlinear preferential attachment mechanism (Eq.\ref{rate}) with similar parameters $\alpha \approx 1.2-1.3$, $ k_{0} \approx 1$ and the aging function decaying with time as $A\propto t^{-1.6}$ in the first decade after publication. (Divergence of $A$ after 10 years is most probably related to the different annual growth rate of the number of publications in these disciplines.) The  Pearson autocorrelation coefficient for the Mathematical and Economics papers (not shown here)  can be approximated by the same dependence, $c_{t,t-1}=(k+3)/(k+60)$, found for the Physics papers. All this provides strong evidence of the generality of our results. 
%, and by the different annual growth rate  the number of Physics papers grew with the annual  rate of $2\%$ while the number of Math papers grew with the annual rate of $4\%$. 
%A very similar parameters were found in the studies of US patens citations \cite{Csardi}. 


\begin{thebibliography}{}

%complex networks
\bibitem{Barabasi} R. Albert and A.L. Barabasi, Rev. Mod. Phys.  \textbf{74}, 47 (2002).
\bibitem{Dorogovtsev2002} S.N. Dorogovtsev and J.F.F. Mendes, Adv. Phys. 
\textbf{51}, 1079 (2002).
\bibitem{Newman_SIAM}M. E. J. Newman, SIAM review, \textbf{45}, 167 (2003); Physics Today, 33, November (2008).

\bibitem{Simon} H.A. Simon, Biometrika \textbf{42}, 425 (1955).
\bibitem{Solla} D. de Solla Price, Science  \textbf{149}, 510 (1965); J.  Am.  Soc. for Inform.  Science  \textbf{27}, 292 (1976).

\bibitem{KRL} P. L. Krapivsky, S. Redner, and F. Leyvraz, Phys. Rev. Lett. \textbf{85}, 4629 (2000).
\bibitem{Burrell}Q.L. Burrell, J. Am. Soc. for Inform. Science, \textbf{54}, 372 (2003).



\bibitem{Jeong} H. Jeong, Z. Neda, and A.L. Barabasi, Europhys. Lett.  \textbf{61}, 567 (2003).
\bibitem{Eom}Y.-H. Eom, C. Jeon, H. Jeong, and B. Kahng, Phys. Rev. E \textbf{77}, 056105 (2008).
%\bibitem{Eom1}Y.-H. Eom and S. Fortunato, PLoS ONE \textbf{9}, e24926 (2011).
\bibitem{Redner2005} S. Redner, Physics Today \textbf{58}, 49 (2005).
\bibitem{Wang2008} M. Wang, G. Yu, and D. Yu, Physica A \textbf{387}, 4692  (2008). 

\bibitem{Borner} A. Scharnhorst, K. Borner, and  P. van den Besselaar, \emph{Models of Science Dynamics} (Springer, Berlin, 2012).


\bibitem{Bianconi} G. Bianconi and A.L.Barabasi, Phys. Rev. Lett. \textbf{86}, 5632 (2001).
\bibitem{Medo}M. Medo, G. Cimini, and S. Gualdi, Phys. Rev. Lett. \textbf{107}, 238701 (2011).

\bibitem{Valverde} S. Valverde, R.V.Sole, M.A.Bedau, and N. Packard, Phys. Rev. E.\textbf{76}, 056118 (2007).



\bibitem{Csardi} G. Csardi, K.J. Strandburg, L. Zalanyi, J.Tobochnik, and P. Erdi, Physica A  \textbf{374}, 783 (2007).


\bibitem{Hawkes}A. G. Hawkes, Biometrika  \textbf{58}, 83 (1971). 
\bibitem{Cattuto} C. Cattuto, V. Loreto, and V.D.P.Servedio, Europhys. Lett.  \textbf{76}, 208 (2006).
\bibitem{mover} M. E. J. Newman, Europhys. Lett. \textbf{86}, 68001 (2009).

%power-law
\bibitem{Redner1998} S. Redner, Eur. Phys. J. B  \textbf{4}, 131 (1998). 
\bibitem{Wallace} M.L. Wallace, V. Lariviere, and Y. Gingras, J. Informetrics  \textbf{3}, 296 (2009). 
\bibitem{Peterson} G.J. Peterson, S. Presse, and K.A. Dill, Proc. Natl. Acad. Sci. USA \textbf{107}, 16023 (2010).


%log-normal
\bibitem{Petersen} A.M. Petersen, F. Wang, and H.E. Stanley, Phys. Rev. E \textbf{81}, 036114  (2010).
\bibitem{Amaral} M.J. Stringer, M. Sales-Pardo, and L.A.N. Amaral, PLoS ONE  \textbf{3}, e1683, (2008).  
\bibitem{Fortunato} F. Radicchi, S. Fortunato, and C. Castellano,  Proc. Natl. Acad. Sci. USA  \textbf{105}, 17268 (2008).

\bibitem{Dorogovtsev-aging}S. N. Dorogovtsev and J. F. F. Mendes, Phys. Rev. E  \textbf{62}, 1842 (2000).

%single refernces
\bibitem{Golosovsky}M. Golosovsky and S.Solomon, Eur. Phys. J \textbf{205}, 303 (2012).
\bibitem{Goffman} W. Goffman and V.A. Newill, Nature \textbf{204}, 225 (1964).
\bibitem{Jaffe} A.B. Jaffe and M. Trajtenberg, Proc. Natl. Acad. Sci. USA \textbf{93}, 12671 (1996).
\bibitem{Bradford} S.C. Bradford,  Engineering  \textbf{137}, 85 (1934). Reprint:  Journal of Information Science, \textbf{10}, 176 (1985).
\bibitem{Fortunato} F. Radicchi, S. Fortunato, and C. Castellano,  Proc. Natl. Acad. Sci. USA  \textbf{105}, 17268 (2008).

\end{thebibliography}
\end{document}